\documentclass[aps,twocolumn,show:pacs,prl,floatfix]{revtex4}
\usepackage{graphicx}
\usepackage{epsf}
\usepackage{wrapfig}
\usepackage{epsfig}
\usepackage{sublabel}
\usepackage{epsfig}
\def\Vec#1{\mbox{\boldmath $#1$}}
\def\bra#1{\langle#1|}
\def\ket#1{|#1\rangle}
\def\beq{\begin{equation}}
\def\eeq{\end{equation}}

\def\beqy{\begin{eqnarray}}
\def\eeqy{\end{eqnarray}}

 \advance\oddsidemargin by -1in
 \advance\evensidemargin
 by -1in
 \marginparsep
 0.4cm
 \advance\topmargin by
 -0.0in
 \makeindex
\def\Vec#1{\mbox{\boldmath $#1$}}
\def\bra#1{\langle#1|}
\def\ket#1{|#1\rangle}

\def\beq{\begin{equation}}
\def\eeq{\end{equation}}

\def\beqy{\begin{eqnarray}}
\def\eeqy{\end{eqnarray}}

\newcommand{\ber}{\begin{displaymath}}
\newcommand{\eer}{\end{displaymath}}
\newcommand{\bey}{\begin{eqnarray}}
\newcommand{\eey}{\end{eqnarray}}

\def\Vec#1{\mbox{\boldmath $#1$}}
\def\bra#1{\langle#1|}
\def\ket#1{|#1\rangle}
\def\beq{\begin{equation}}
\def\eeq{\end{equation}}

\def\beqy{\begin{eqnarray}}
\def\eeqy{\end{eqnarray}}

\begin{document}
\vskip 2mm \date{\today}\vskip 2mm
\title{Effects of Ground-State Correlations on High Energy Scattering off Nuclei: the
Case of the
 Total Neutron-Nucleus Cross Section}
\author{M. Alvioli}
\author{C. Ciofi degli Atti}
\author{I. Marchino}
\author{V. Palli}\affiliation{Department of Physics, University of Perugia and
      Istituto Nazionale di Fisica Nucleare, Sezione di Perugia,
      Via A. Pascoli, I-06123, Italy}
\author{H. Morita}\affiliation{Sapporo Gakuin University, Bunkyo-dai 11, Ebetsu 069,
  Hokkaido, Japan}
\pacs{21.30.Fe, 21.60.-n, 24.10.Cn, 25.30.-c}
\begin{abstract}
\noindent With the aim at quantitatively investigating the
longstanding problem concerning the effect of short range nucleon-nucleon
correlations on scattering processes at high energies, the total
neutron-nucleus cross section is calculated within a
parameter-free approach which, for the first time,   takes into
account,  simultaneously,  central, spin, isospin and tensor
 nucleon-nucleon (NN) correlations, and  Glauber elastic  and
Gribov inelastic shadowing corrections. Nuclei  ranging
from $^{4}He$ to $^{208}Pb$  and   incident neutron momenta
in the range  $3$ $GeV/c$ - $300$ $GeV/c$ are  considered; the commonly used
approach which approximates  the square of the nuclear wave
function by a product of one-body densities is carefully
analyzed, showing that NN correlations can play a non-negligible
role in high energy scattering off nuclei
\end{abstract}
\maketitle
\begin{figure*}[!htp]
  \centerline{\epsfig{file=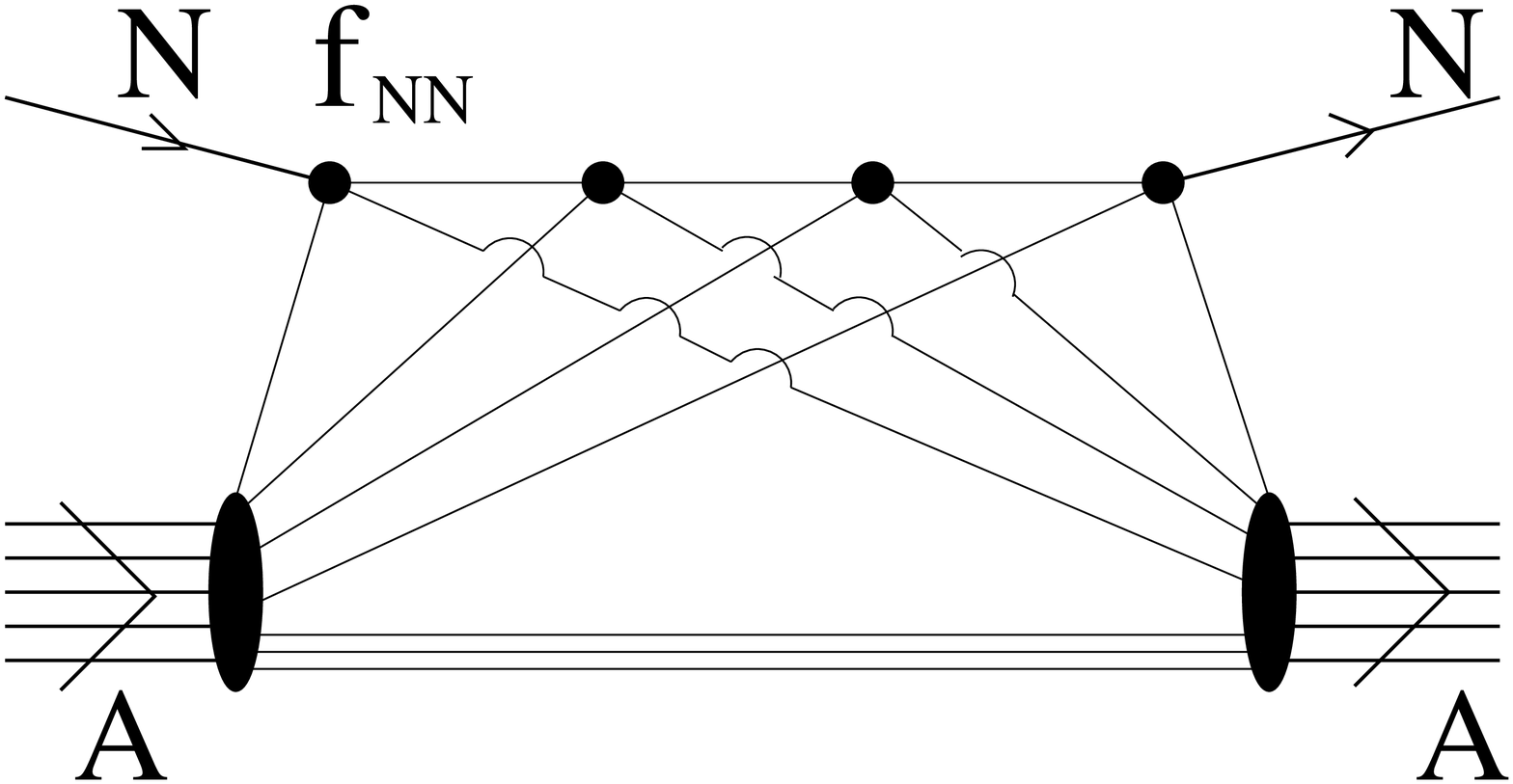,width=4.5cm}
    \hspace{0.5cm}\epsfig{file=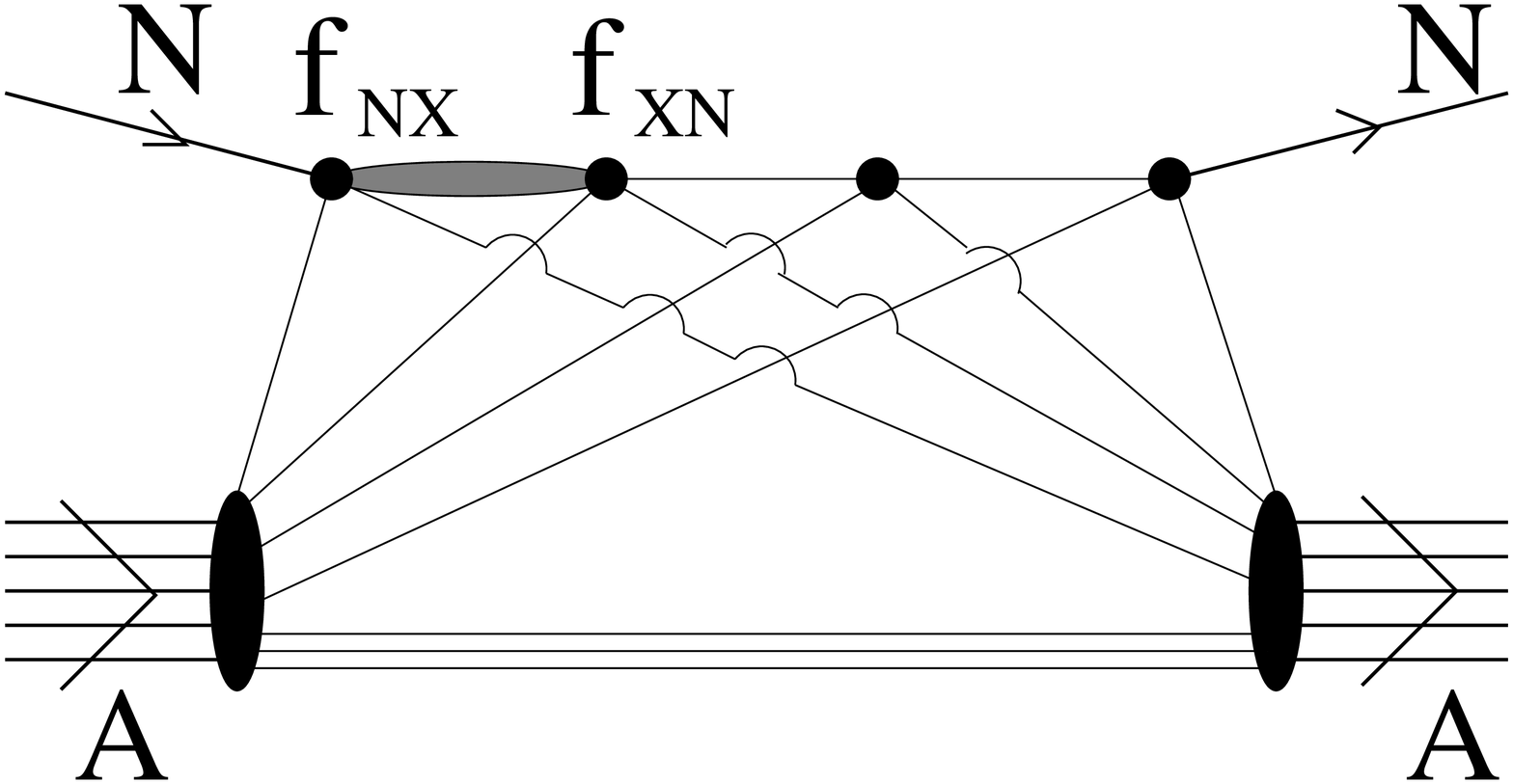,width=4.5cm}
    \hspace{0.5cm}\epsfig{file=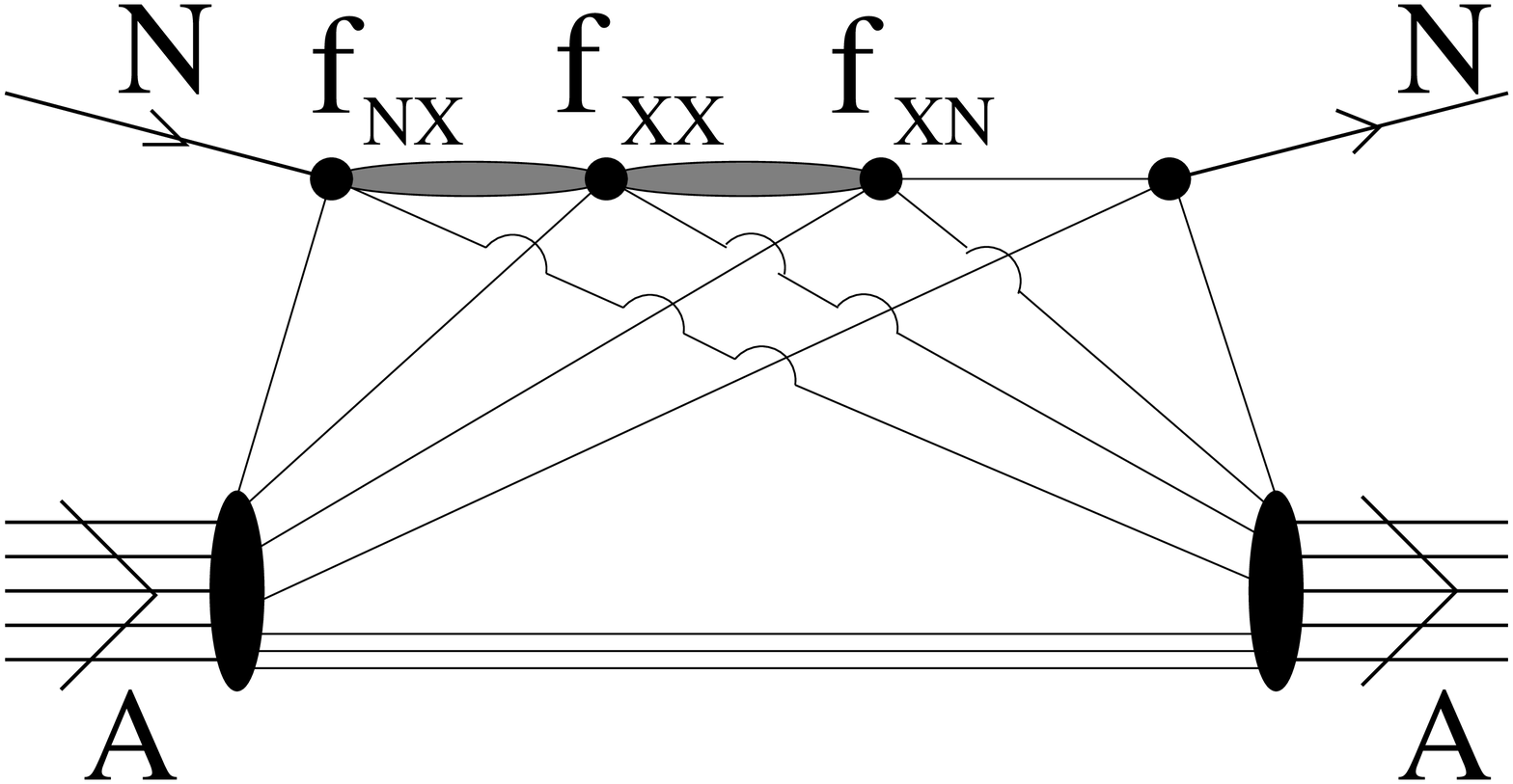,width=4.5cm}}
\centerline{\large a)\hspace{4.8cm}b)\hspace{4.8cm}c)\normalsize}
  \caption{Typical diagrams describing elastic N-A scattering:
   \textit{a)} Glauber elastic scattering; \textit{b)}
    and \textit{c)} Gribov inelastic scattering. Black dots denote the particle-particle scattering
    amplitude.}\label{Fig1}
\vskip -0.4cm
\end{figure*}

\begin{figure}[!hbp]
\vskip -0.5cm
\centerline{\centerline{\epsfig{file=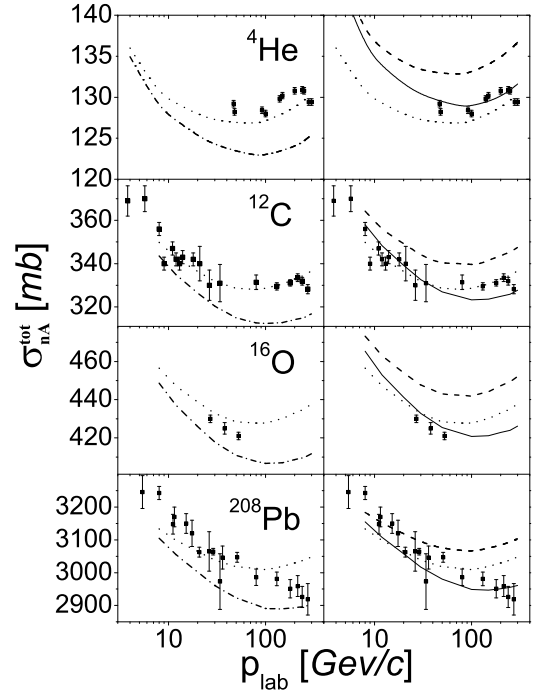,width=8cm}}} \vskip
-0.3cm \caption{$\sigma_{tot}^{nA}$ {\it vs} $p_{lab}$. \textit{Left panel}:
  Glauber single density approximation ($\sigma_G$; \textit{dots})
  and Glauber plus Gribov inelastic shadowing
  ($\sigma_G$ + $\Delta\sigma_{IS}$; \textit{dot-dash}).
  \textit{Right panel}:
  Glauber ($\sigma_G$; \textit{dots});
  Glauber   plus SRC ($\sigma_G$ + $\sigma_{SRC}$; \textit{dashes});
  Glauber   plus SRC
  plus Gribov inelastic shadowing  ($\sigma_G$ + $\sigma_{SRC}$ + $\Delta\sigma_{IS}$;
    \textit{full}).
  Experimental data from
  \cite{murthy,data02}.}\label{Fig2}
\end{figure}

 Nowadays interpretation of high precision
particle-nucleus and nucleus-nucleus scattering experiments at
medium and high  energies, aimed at investigating the state of
matter at short distances, should require in principle also a
consideration  of possible effects from short range NN correlations (SRC),
 particularly  in
view of   recent experimental data on lepton and hadron
scattering off nuclei which provided  quantitative
evidence  on SRC and their  possible effects on dense  hadronic matter
\cite{eli}. Thanks to recent progress in the theoretical
description of the many-body nuclear wave function, we have therefore
undertaken a systematic study of the effects of SRC in
medium and high  scattering of nuclei  starting with a novel calculation of
the total neutron-nucleus cross section $\sigma^{tot}_{nA}$ at
high energies. This quantity has been experimentally measured with
high precision in a wide kinematical range and has been the object
of many theoretical analyzes since it appears  to be  very
sensitive to various relevant phenomena, such as Glauber elastic
\cite{glauber} and Gribov  inelastic \cite{gribov} diffractive
shadowing, which, in turn, have a relevant  impact on the
interpretation of color transparency phenomena
and  relativistic heavy ion processes (see e.g. \cite{jen01,boris2} )
It is well known that although the major mechanism which explains
the experimental evidence  $\sigma^{tot}_{nA} <<
A\,\sigma_{N}$ ($\sigma_{N}\equiv \sigma^{tot}_{NN}$) is Glauber
elastic shadowing, a quantitative explanation of the experimental
data requires also the introduction of Gribov inelastic shadowing
\cite{jen01,murthy,nikolaev}.
 Most  calculations of $\sigma^{tot}_{nA}$ so far performed were however based upon the
so called one-body-density approximation, in which all terms but
the first one of the exact  expansion of the square of the nuclear
wave function in terms of density matrices \cite{glauber,foldy}
are disregarded, which amounts to neglect all kinds of NN
correlations.  Although  the necessity and interest to investigate
the effects of the latter have been stressed  by several authors
\cite{jen01,nikolaev}, first of all by Glauber himself
\cite{glauber},  only few qualitative calculations have  been performed
\cite{moniz,rafa}. The aim of this work is to illustrate  a novel
parameter-free  calculation of $\sigma^{tot}_{nA}$ within a
realistic treatment of SRC \cite{alv01,alv02}. In terms
of
 Glauber (G) elastic and Gribov
inelastic (IS) scattering one has
\beq
\sigma_{nA}^{tot}\,=\,\sigma_A^G\,+\,{\sigma}_A^{IS}\,=\,\frac{4\pi}{k}\,Im
\left[\,F_{00}^G(0)\,+\,F_{00}^{IS}(0)\,\right]
\eeq
where $F_{00}^{G(IS)}(0)=\frac{ik}{2\pi}\,\int
d\Vec{b}_n\,\Gamma_{00}^{G(IS)}(\Vec{b}_n)$ denotes the forward
elastic scattering amplitude, and $\Gamma_{00}^{G(IS)}$ the nuclear elastic
profile function, namely
\begin{equation}
\Gamma_{00}^G(\Vec{b}_n)\,=\,1\,-\,\prod_{j=1}^A\langle \psi_0
\left |\left[\,1\,-\,\Gamma_{N}(\Vec{b}_n-{\bf s}_j)\,\right]
\right | \psi_0\rangle\,, \label{SG}
\end{equation}
Here $\psi_0 \equiv \psi_0({\bf r}_1,{\bf r}_2,{\bf r}_3,...{\bf
r}_A)$, with ${\bf r}_j=({\bf  s}_j,z_j)$, is  the ground state wave
function of the target nucleus, $\Vec{b}_n$  the impact parameter
of the neutron moving along the $z$-axis, and
${\mit\Gamma}_{N}(\Vec{b}_n)$ the NN elastic  profile function. As
for the Gribov inelastic  profile, it describes, as depicted in
Fig.\ref{Fig1}, the diffractive dissociation of the neutron via
the process $n + N \rightarrow X + N$,  its  de-excitation to the
ground state by the process $X + N \rightarrow n + N$, and its
 elastic scattering  off the target nucleons. In our approach,
as in Ref. \cite{jen01}, we will consider, besides   the elastic
scattering of $X$, only two non-diagonal transitions ($n + N
\rightarrow X + N$ and $X + N \rightarrow n + N$).
 Within such an
approximation one obtains  \cite{jen01}:
\begin{widetext}
\beqy
\Gamma_{00}^{IS}(\Vec{b}_n)&=&
\sum_{X}\left\{\bra{\psi_o}\sum_{i<j}^{A}\,\Gamma_{NX}(\Vec{b}_n-\Vec{b}_j)
\,\Gamma_{XN}(\Vec{b}_n-\Vec{b}_{i})\,e^{i\,q_{X}(z_i-z_j)}\Theta (z_j-z_i)
\,\times\right.\nonumber\\
&&\hspace{-1.5cm}\left.\times \prod_{k\neq i,j}^{A}[1-\Gamma_{X}(\Vec{b}_n-\Vec{b}_{k})]
\,\Theta(z_k-z_i)\,\Theta(z_j-z_k)
\,\prod_{l\neq i,j}^{A}[1-\Gamma_{N}(\Vec{b}_n-\Vec{b}_{l})]
\,\Theta(z_i-z_l)\,\Theta(z_l-z_j)\ket{\psi_o}\right\}\,,
\label{Fdelta}
\eeqy\end{widetext}
where \,$q_{X}\,=\,k_{n}\,-\,k_{X}$ is the longitudinal momentum
transfer. The basic nuclear ingredient appearing in Eqs.
(\ref{SG}) and (\ref{Fdelta}) is $|\psi_0|^2$, which, in terms of
density matrices, has the following form  \cite{glauber,foldy}
(the Center-of-Mass $\delta$ function is omitted for ease of
presentation):
\begin{widetext}
\beq
\left|\,\psi_o(\Vec{r}_1,...,\Vec{r}_A)\,\right|^2=\prod_{j=1}^A\,\rho_1(\Vec{r}_j)
\,+\,\sum_{i<j}\,\Delta(\Vec{r}_i,\Vec{r}_j)\hspace{-0.1cm}\prod_{k\neq i,j}\rho_1(\Vec{r}_k)\,+
\hspace{-0.5cm}\sum_{(i<j)\neq(k<l)}\hspace{-0.5cm}\Delta(\Vec{r}_i,\Vec{r}_j)
\,\Delta(\Vec{r}_k,\Vec{r}_l)\hspace{-0.3cm}\prod_{m\neq i,j,k,l}\hspace{-0.3cm}\rho_1(\Vec{r}_m)
\,+\,\dots\,,
\label{psiquadro}
\eeq
\end{widetext}
in which $\rho_{1}(\Vec{r}_i)$ is the one-body density matrix
(normalized to one) and
${\Delta(\Vec{r}_i,\Vec{r}_j)}\,=\,\rho_2(\Vec{r}_i,
\Vec{r}_j)\,-\,\rho_{1}(\Vec{r}_i)\,\rho_{1}(\Vec{r}_j)$
 the
\textit{two-body contraction}. Here the two-body density
matrix $\rho_2(\Vec{r}_i,\Vec{r}_j)$ must
satisfy  the sequential condition
$\int\,d\Vec{r}_j\,\rho_2(\Vec{r}_i,\Vec{r}_j)\,=\,\rho_1(\Vec{r}_i)$
 leading to $\int d\Vec{r}_j\,\Delta(\Vec{r}_i,\Vec{r}_j)\,=0$.
Note that in  Eq. (\ref{psiquadro}) only   unlinked contractions
have to be considered, and that the  higher order terms, not
explicitly displaced,  include unlinked products of 3, 4, {\it
etc} two-body contractions, unlinked products of three-body
contractions, describing three-nucleon correlations, and so on. By
taking into account two-body correlations only, i.e.  all terms
of the expansion (\ref{psiquadro}) containing all possible numbers
of unlinked two-body contractions, one obtains \cite{moniz,rafa} (
from now-on the optical limit, $A>>1$ will be used for ease of
presentation):
\begin{widetext}
\beq \Gamma_{00}^G(\Vec{b}_n)\simeq 1-\exp\left[ -A\int d\Vec{r}_1
      \,\rho_{1}(\Vec{r}_1)\,\Gamma(\Vec{b}_n-\Vec{s}_1)+
\frac{A^2}{2}{\int d\Vec{r}_1 d\Vec{r}_2\,\Delta(\Vec{r}_1,\Vec{r}_2)\,
      \Gamma(\Vec{b}_n-\Vec{s}_1)\,\Gamma(\Vec{b}_n-\Vec{s}_2)}\right]
\label{optical}
\eeq
\end{widetext}
which yields the usual Glauber profile when $\Delta = 0$.
Concerning $\Gamma_{00}^{IS}$,
it can be reduced to an expression depending upon the total
nucleon and diffractive cross sections $\sigma _N$ and $\sigma _r$
respectively \cite{jen01} which, within the
approximation $\sigma _N= \sigma _r$ and disregarding
correlations, provides the well-known Karmanov-Kondratyuk (KK)
result \cite{karmanov} :
\begin{widetext}
\beq
\Gamma_{00}^{IS}({\bf b}_n)\,=\,-(2\pi)A^2\,\int
\frac{d^2\sigma}{d^2q_{T}\,dM_{X}^2}\,\Big|_{q_{T}=0} dM_{X}^2
e^{\displaystyle{-\frac{\sigma_{N}}{2}\,T({\bf b}_n)}}|F(q_{L}, {\bf b}_n)|^2\,,
\label{KK}
\eeq
\end{widetext}
Here $T({\bf b}_n)= A\int_{-\infty}^{\infty}\rho({\bf b}_n,z)\,dz$
is the thickness function, $F(q_{L},{\bf b}_n)=
\int_{-\infty}^{\infty}\rho({\bf b}_n,z)\,exp(iq_L\,z)\,dz$ is the
nuclear form factor, depending upon $M_X$ through the relation
$q_L=(M_X^2-m_N^2)m_N/s$,
 and $d^2\sigma/(d^2q_{T}\,dM_{X}^2)$ is the differential
cross section of the process $N+N \rightarrow N_X +N$ ($M_X$ being
the mass of $N_X$). We have calculated $\sigma^{tot}_{nA}$ using
the two-body density  obtained from the  fully-correlated wave
function of Ref. \cite{alv01,alv02}, $\psi_0 ={\hat F} \phi_0$,
where ${\hat F} =\prod_{i=1}^8 {\hat f}_{ij}$ is  a correlation
operator generated by the realistic Argonne $V8^\prime$
interaction \cite{vuotto}, and $\phi_0$ a mean field (MF) wave
function. The above wave function largely differs from the Jastrow
wave function, featuring  only central correlations, since the
operator ${\hat F}$  generates central, spin, isospin, tensor,
etc. correlations.  The one-body density has been obtained by
integrating the two-body density, and the contraction
$\Delta(\Vec{r}_1,\Vec{r}_2)$,  exactly satisfying the sequential relation,
has been obtained
(note that our one-body point density and radii are in agreement
with electron scattering data \cite{electron}).
The Glauber
profile has been chosen in the usual form,
$\Gamma(\Vec{b}_n)\,=\,{\sigma_{tot}}{(4\,\pi\,b_0^2)}^{-1}\,(1-i
\alpha)\,exp{(\displaystyle{-\Vec{b}_n^2 / 2\,b_0^2})}$, with the
energy-dependent parameters taken from \cite{param_glau}; the
parameters for the inelastic shadowing were taken from
\cite{murthy}. The results of calculations for $^4He$, $^{12}C$ ,
$^{16}O$ and $^{208}Pb$ are presented in Fig. \ref{Fig2}. The left
panel shows the results obtained without correlations, i.e. taking
into account only the first term in the exponent of Eq.
(\ref{optical}), whereas the results presented in the right panel
include the effects of SRC by considering both terms in
the exponent (for $^4He$ we have calculated the
cross section to all orders  finding  that three- and
four-nucleon correlations produce negligible effects).
The results presented in Fig.2  show that: i)
within the one-body density approximation,
 inelastic shadowing corrections
 {\it increase} the nuclear transparency, which is a well-known result,
 but, at the same
time, if realistic one-body densities are considered, as in the present paper,
they worsen
the agreement with the experimental data;
this result is  at variance with Ref. \cite{murthy} where too large (by about $15 \%$)
 nuclear radii have
been used, as first stressed in \cite{nikolaev};
ii) NN correlations {\it decrease}  the transparency (which is physically due to the reduction of the role
of Glauber shadowing) and
increase the total cross section by an amount ranging from about $2 \%$ in
$^{208}Pb$ up to about $5-6\%$ in $^3He$, spoiling   the
agreement with the experimental data provided by the Glauber
calculation ; iii) the inclusion of
inelastic shadowing brings back theoretical calculations in good
agreement with experimental data. Thus
it appears that if the correct values of  nuclear radii are used,
the interpretation of the experimental data would
require the consideration of {\it both} NN correlations and
inelastic shadowing.
We have also investigated the validity of the approximation
consisting in using  for finite nuclei the nuclear matter
 two-body density, $viz$
$\rho_2(\Vec{r}_1,\Vec{r}_2)\,=\,\rho_1(\Vec{r}_1)\,\rho_1(\Vec{r}_2)
\,g(|\Vec{r}_1-\Vec{r}_2|)$
which, for nuclei with $A < 208$, strongly violates
the sequential relation
 $\int d {\bf r}_2
\rho_2(\Vec{r}_1,\Vec{r}_2)= \rho_1(\Vec{r}_1)$,
which means that using it  to introduce
correlations in light and  medium-weight nuclei generates a
mismatch between the one-body density (usually taken from the
experimental data) and the two-body density.
\begin{figure}[!htp]
  \vskip -0.5cm \centerline{
    \epsfysize=0.4\textwidth\epsfbox{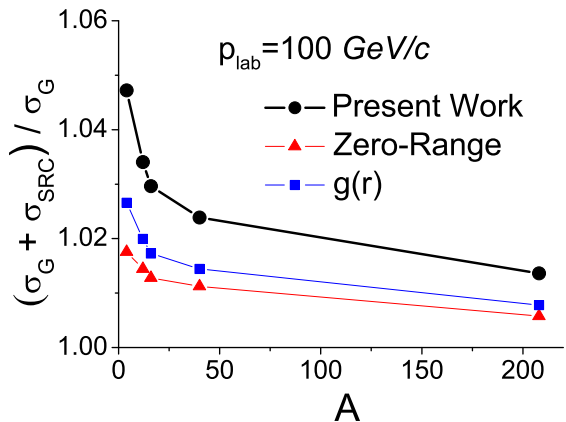}}
  \vskip -0.3cm \caption{Ratio of the total cross section which includes correlations
  to the cross section without correlations;  the figure shows  our  results
obtained using the  finite nuclei ({\it dots}) and nuclear  matter (${\it squares}$)
two-body
  densities. {\it Triangles} correspond to the zero range approximation
  for the profile function.}
  \label{Fig3}
\end{figure}

To sum up, we have analyzed the effects of SRC on
$\sigma_{tot}^{nA}$ within a realistic and parameter-free description
of SRC,  using  the correct values of nuclear
radii and, at the same time,
one-body  densities  which, unlike previous calculations, are exactly
linked  to the two-body densities by  the sequential relation. The results
we have obtained
show that the effects of SRC, though being small in  absolute value,
could be of the same order as
Gribov inelastic shadowing corrections.
Such a result  points to the necessity of: i) a systematic investigation of
SRC effects on other high energy scattering processes (e.g. electro-production of
hadrons, large rapidity gap processes \cite{boris2}, heavy-ion
collisions \cite{mark3}, etc);  ii) an improved treatment
 of  Gribov inelastic shadowing, going beyond the  lowest order
intermediate diffractive excitations. To conclude, we would like to point
 out that the
smallness of SRC effects on $\sigma_{tot}^{nA}$ does not imply  that SRC effects on
other quantities will also be small; as a matter of fact,
 preliminary results \cite{MCV} show that SRC reduce the  quasi-elastic
 cross section $\sigma_{qel}^{pA}$ up to $15 \%$ in $^{12}C$ and $^{208}Pb$.
 Calculations of elastic and quasi-elastic cross sections
 at energies ranging from HERA to LHC, are in progress and will be reported elsewhere.

CdA would like to thank B. Kopeliovich, M. Strikman and D. Treleani  for many useful discussions and
N. Nikolaev for his suggestion to investigate  the effects of SRC on the total cross
 sections. This work was supported in part by Fondecyt (Chile) grant 7070223; CdA is
  grateful to  Departamento de Fisica, Universidad
Tecnica Federico Santa Maria, Valparaiso, Chile, where the manuscript has been completed.
\setcounter{footnote}{0}






\end{document}